# Air, bone and soft-tissue Segmentation on 3D brain MRI Using Semantic Classification Random Forest with Auto-Context Model

Xue Dong[a#], Yang Lei[a#], Sibo Tian[a], Yingzi Liu[a], Tonghe Wang[a], Tian Liu[a],
Walter J. Curran[a], Hui Mao[b], Hui-Kuo Shu[a] and Xiaofeng Yang[a*]

[a] Department of Radiation Oncology and Winship Cancer Institute, Emory University, Atlanta, GA 30322

[b] Department of Radiology and Imaging Sciences and Winship Cancer Institute, Emory University, Atlanta, GA 30322.

# Equal contribution, co-first author

*Corresponding to: xiaofeng.yang@emory.edu

**Abstract**

As bone and air produce weak signals with conventional MR sequences, segmentation of these tissues particularly difficult in MRI. We propose to integrate patch-based anatomical signatures and an auto-context model into a machine learning framework to iteratively segment MRI into air, bone and soft tissue. The proposed semantic classification random forest (SCRF) method consists of a training stage and a segmentation stage. During training stage, patch-based anatomical features were extracted from registered MRI-CT training images, and the most informative features were identified to train a series of classification forests with auto-context model. During segmentation stage, we extracted selected features from MRI and fed them into the well-trained forests for MRI segmentation. The DSC for air, bone and soft tissue obtained with proposed SCRF were 0.976±0.007, 0.819±0.050 and 0.932±0.031, compared to 0.916±0.099, 0.673±0.151 and 0.830±0.083 with RF, 0.942±0.086, 0.791±0.046 and 0.917±0.033 with U-Net. SCRF also demonstrated superior segmentation performances for sensitivity and specificity over RF and U-Net for all three structure types. The proposed segmentation technique could be a useful tool to segment bone, air and soft tissue, and have the potential to be applied to attenuation correction of PET/MRI system, MRI-only radiation treatment planning and MR-guided focused ultrasound surgery.

# 1. Introduction

Magnetic resonance image (MRI) is a widely accepted modality for cancer diagnosis and radiotherapy target delineation due to its superior soft tissue contrast. Bone and air segmentations are important tasks for MRI, and facilitate several clinical applications, such as MRI-based treatment planning in radiation oncology[1], MRI-based attenuation correction for positron emission tomography (PET) [2], and MR-guided focused ultrasound surgery (FUS) [3].

A treatment planning process with MRI as the sole imaging modality could eliminate systematic MRI-CT co-registration errors, reduce medical cost, spare patients from CT x-ray exposure, and simplify clinical workflow. However, MRI data do not contain the electron density information that is necessary for accurate dose calculation and generating reference images for patient setup [4-9]. While ignoring inhomogeneity gives rise to 4-5% of dose errors, simply assigning three bulk densities, such as, bone, tissue and air, could reduce deviations to less than 2% , which is clinically acceptable [10]. Even with the use of synthetic CTs that provides continuous electron density estimation for improved dose calculation accuracy, bone and air identification is considered key to heterogeneity correction and thus accurate dose estimation [11]. The hybrid PET/MRI system has emerged as a promising imaging modality due to the unparalleled soft tissue information provided by the non-ionizing imaging modality. Though different types of MR-based attenuation correction methods have been investigated, virtually all current commercial PET/MRI systems employ segmentation-based methods due to its efficiency, robustness and simplicity [12]. Accurate segmentation of different tissue types, especially of bone and air, directly impacts the estimation accuracy of attenuation map. Bone segmentation in MRI also facilitates the quickly-developing technology of image-guided FUS [3]. FUS requires a refocusing of the ultrasound beams to compensate for distortion and translation caused by the attenuation and scattering of the beams through bone [13]. Since many procedures utilize, and are based on MR capabilities, it would be desirable to delineate bone from MRI, potentially avoiding the additional steps of CT acquisition and subsequent CT-MRI co-registration.

In contrast to CT, which provides excellent tissue-bone and tissue-air contrast, both bone and air are of low proton density and produce weak signals with conventional MR sequences, making bone and air segmentation particularly difficult. To accurately delineate soft tissue, air and bone, a straightforward approach is to warp atlas templates to the MRI, allowing one to exploit the excellent bone and air contrast on CT images to identify the corresponding structures in MRI [14, 15]. Besides computational cost, atlas-based methods are prone to registration errors as well as inter-patient variability. A larger and more varied atlas dataset could help to improve registration accuracy. However, because of organ morphology and substantial variability across patients, it is difficult to satisfy all possible scenarios. Moreover, larger atlas templates are usually associated with significantly increased computational cost. Specialized MR sequences, such as ultrashort echo time (UTE) pulse sequences, have been investigated for bone visualization and segmentation. Its performance is limited by of noise and image artefacts [16, 17]. Moreover, due to considerable long acquisition time, the application of UTE MR sequences is usually limited to brain imaging or small field-of-view images. Machine learning-based and deep learning-based segmentation and synthetic CT generation methods have been intensively studied for the last decades[18-30], among which random forest-based method is one popular machine learning approach. The popularity of random forest arises from its appealing features, such as its capability of handling a large variety of features and enabling feature sharing of a multi-class classifier, robustness to noise and efficient parallel processing [31]. Random forest has been employed to generate synthetic MRIs of different sequences for improved contrast [32], synthetic CTs for MRI-only radiotherapy treatment planning [18, 19], as well as PET AC [33].

In this work, we propose to integrate an auto-context model and patch-based anatomical signature into a random forest framework to iteratively segment air, soft tissue and bone on routine anatomical MRIs.

This semantic classification random forest (SCRF)-based approach has 3 distinctive strengths: 1) In order to enhance feature sensitivities to detect structures, three types of features are chosen to characterize information of an image patch at different levels from voxel level, sub-region level, to whole-patch level. 2) In order to improve the random forest training efficiency, a feature selection mechanism is introduced to identify the most informative and salient features from the extracted features through minimizing the logistic sparse LASSO energy function. The selected features with higher discriminative power are used to train the random forest. 3) Contrary to the traditional segmentation or classification methods, an auto-context model is used to incorporate the context information from the previously discriminative probability maps in random forest framework to provide an iterative refinement for the final segmentation, which significantly improves the segmentation accuracy. To demonstrate the effectiveness of the auto-context model [25], we compared the performance of the proposed method with a conventional random forest framework without the auto-context model. Deep learning-based methods show state-of-the-art performances in various medical imaging applications. Therefore, we trained a well-established deep learning-based model, U-Net[34], and compared its segmentation accuracy with the proposed method.

## 2. Methods

*2.1 Method overview*

The proposed method was trained with registered MR and CT images. Given a pair of brain MR and CT training images, the air, soft-tissue and bone labels obtained from CT images were used as the classification target of the MR image. Prior to the training stage, noise regions of the MR image were removed by non-local means method to improve the training quality[35]. The nonparametric nonuniform intensity normalization (N3) algorithm was applied for MR image inhomogeneity bias correction. The intra-subject registration was then performed to align each pair of MR and CT images, as well as the corresponding labels. All pairs were aligned onto a common space by applying rigid-body inter-subject registration. This registration was performed by commercial software, Velocity AI 3.2.1 (Varian Medical

Systems, Palo Alto, CA) using rigid registration. In the training stage, air, soft-tissue, bone labels, and named CT segmentation labels, were clustered from CT images by a fuzzy C-means method. The input patch size of the MRI was [33, 33, 33]. The corresponding CT segmentation label on that patch's central position was regarded as the learning-based classification target. Multi-level features were extracted on voxel, sub-region, and whole-patch levels from each MR image, i.e. pairwise voxel difference, local binary pattern (LBP), and discrete cosine transform (DCT) features from multiscale images, which consisted of the original and 3 derived images with a sequence of down-sampling factors (0.75, 0.5 and 0.25). These extracted features were concatenated as a patch-based feature vector. We then identified the most salient and informative features using a logistic LASSO-based method, a feature selection strategy (as previously recommended [36]) and utilized it together with the corresponding CT segmentation labels to train a sequence of classification forests by integrating an auto-context model [37]. For each classification random forest, we set the number of trees in forest to 100. Minimum Gini impurity optimization was used to create each tree in the forest. A node in a tree will be split if this induces a decrease of the Gini impurity greater than or equal to this value. Gini impurity is a measure of how often a randomly chosen element from the set would be incorrectly labeled if it were randomly labeled according to the distribution of labels in the subset. The Gini impurity can be computed by summing the probability $p_i$ of an item with label $i$ being chosen times the probability $\sum_{k \neq i} p_k = 1 - p_i$ of a mistake in categorizing that item. It reaches its minimum (zero) when all cases in the node fall into a single target category. The weighted impurity decrease equation is calculated as follow:

$$\frac{N_t}{N \cdot \left(I_{Gini} - \frac{N_t^R}{N_t} \cdot I_{Gini}^R - \frac{N_t^L}{N_t} \cdot I_{Gini}^L\right)} \tag{1}$$

where $N$ is the total number of samples, $N_t$ is the number of samples at the current node, $N_t^L$ is the number of samples in the left child node, and $N_t^R$ is the number of samples in the right child node. $I_{Gini}$ denotes the Gini impurity of that set. These classification forests were used to create and improve the segmentation-based context features. By repeating this process until convergence, a sequence of trained forests was obtained. In the segmentation stage, features from the newly acquired MR image were extracted and fed into the trained forests for the segmentation. Fig. 1 outlines the workflow of our segmentation method.

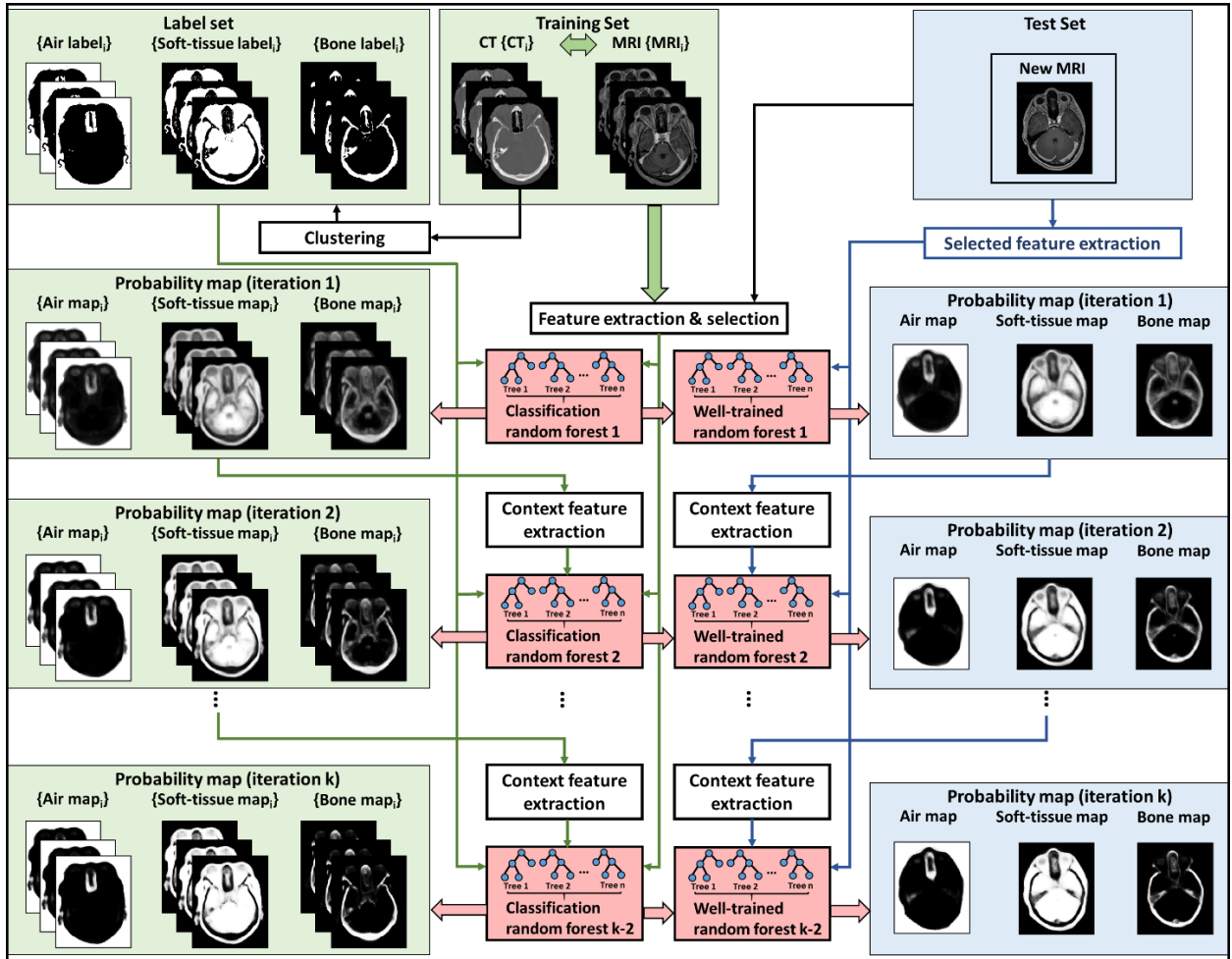

**Fig. 1** The workflow of the proposed brain MRI SCRF method. The training stage is shown on the left, and the segmentation stage is shown on the right.

*2.2 Semantic information*

Traditionally, random forest-based segmentation methods train a classification forest during the training stage and generate the segmentation of newly acquired MRI patch by feeding the features into the trained model. However, MRI patches often have similar intensities or structures around regions that appear differently in CT patches. The extracted features on these MRI patches were similar, which can potentially lead to ambiguity in the model. Classifying similar MR patches to dissimilar CT-based segmentation labels can lead to erroneous classification results. Thus, anatomic features alone (extracted as suggested in our previous work) may not generate accurate tissue segmentations. To cope with this issue, we applied a classification forest to first train a segmentation model for air, bone and soft-tissue labels. Rather than obtaining the binary segmentation results when feeding the features into the trained segmentation model, the posterior probability of air, bone, and soft-tissue under the feeding features can also be estimated by maximizing a posterior. The posterior volume of these three labels are denoted probability maps in this work. The auto-context method was used to iteratively refine the probability maps [38], i.e., the probability maps were updated with each iteration of the classification random forest's training and segmentation. An auto-context method improves segmentation accuracy, since it leverages the information surrounding the object of interest [38]. To better approximate the margins of air, bone, and soft tissue material, we included a set of semantic information, i.e., the surrounding voxels' posterior (probability) under given MRI anatomical features to the objective voxel of interest. The surrounding voxels were located by a number of context locations. Here, we propose to use 27 locations for each voxel of interest. Semantic information was subsequently generated by pairing superior and inferior, left and right, anterior and posterior, and central blocks (with size [3, 3, 3]) within a window (with size [15, 15, 15]) of each probability map. The blocks and window in probability maps are shown in Fig 2.

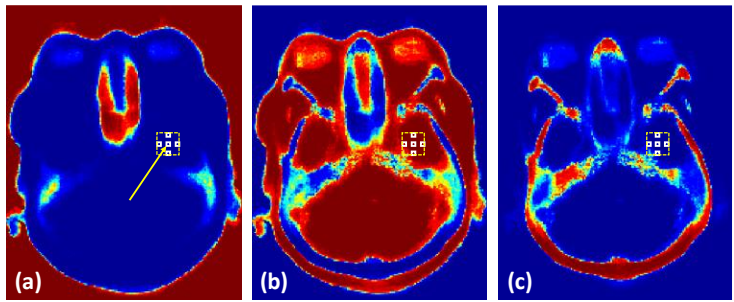

**Fig. 2** An example illustrating the blocks within a window for semantic feature extraction. (a), (b) and (c) show the probability maps of air, soft-tissue and bone. The extracting windows are shown as yellow dotted line rectangles. The extracting blocks are shown as white solid line rectangles. The same display window with [0, 1] is used for all figures.

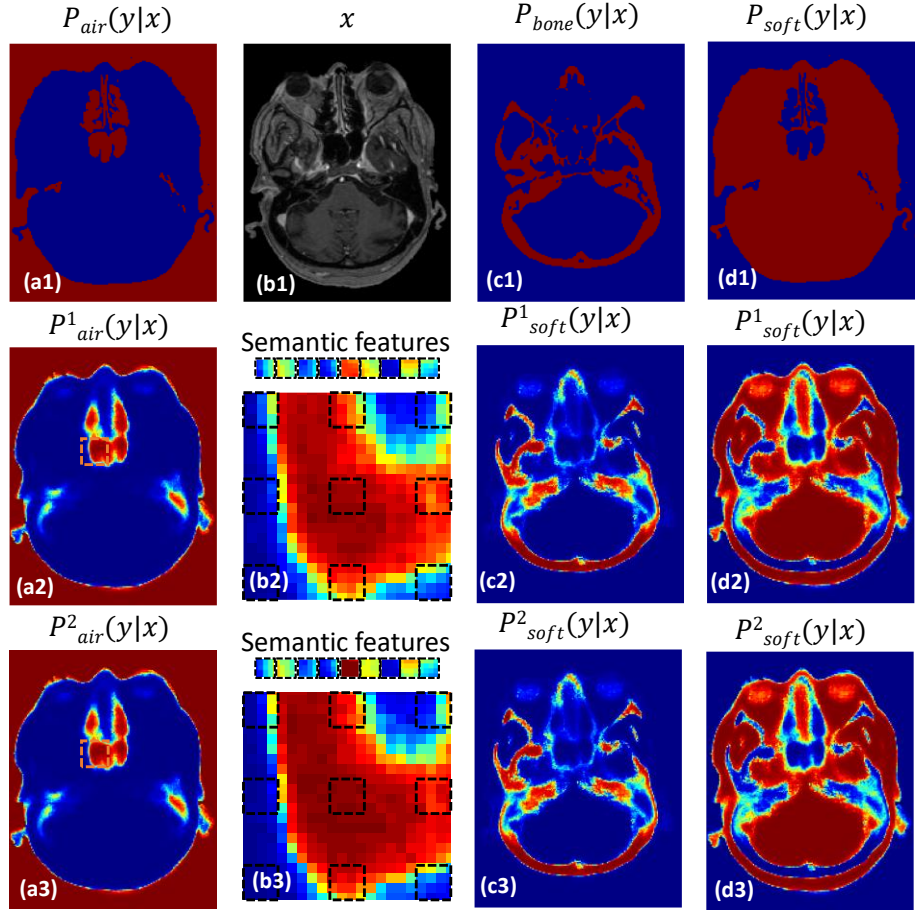

**Fig. 3** An example illustrating the generation of semantic information for incorporation into the random forest model. (b1) is the MRI image, and (a1, c1, d1) show the binary segmentation for each material (air, bone and soft tissue) based on CT segmentation. (a2-a3), (c2-c3), and (d2-d3) show the calculated material probability with increasing iterations. Material probabilities are calculated with a maximum likelihood method conditioned on CT segmentation labels and image features. (b2-b3) show the zoomed in regions near the nasopharynx. The context locations are shown in dotted rectangles of these regions. The semantic information is a concatenation of mean values of these rectangles. Initially, much of the air in this region was calculated to be a mixture of multiple tissue types. As the probability maps are iteratively refined, the classifier correctly determines that many of the pixels in this region are composed of air.

Fig. 3 shows examples of semantic features extracted on context locations from a probability map. Insert (b1) shows an axial MR image, which we denote as *x*. Insert (a1) shows the corresponding air label's probability map, and (a2) shows the air probability map obtained from the first classification random forest-based segmentation model by feeding the features. (b2) shows the semantic feature extraction for the central voxel in the highlighted window of (a2). As shown axially, we see nine highlighted smaller blocks as context locations in (b2). The semantic features were generated by a concatenation of the means within these blocks. Insert (a3) show the air probability map obtained from the second segmentation model. This model was trained not only using the features from the MR image, but also incorporating the semantic features from (b2). Once a new segmentation model was trained, the probability map generation and semantic feature extraction procedure repeats the same procedure until convergence, as is shown in (a4-b4) and (a5-b5). The probability maps of bone and soft tissue are also given in (c1-c3) and (d1-d3). Within the region of interest, the central voxel value of the highlighted window in (a3) has a relatively high posterior for air, while the posterior for bone or soft tissue is low in (c3) and (d3). Thus, this value can clearly separate the image labels, i.e., the separation of materials was best in the fourth segmentation model. The advantage of semantic information is demonstrated in Fig. 4, where semantic features give rise to a better differentiation between air and bone.

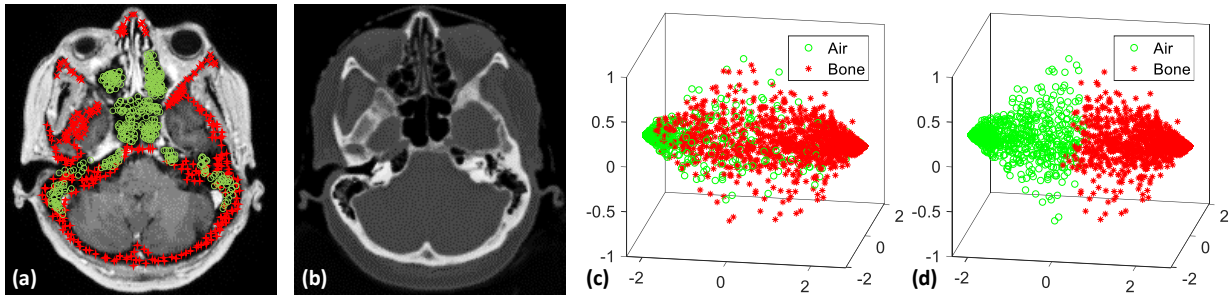

**Fig. 4** An example illustrating the benefit of semantic information. (a) and (b) show axial MRI and CT images, where the samples belonging to the air region are highlighted by green circles, and the samples belonging to bone region are highlighted by red asterisks. (c) shows the scatter plots of first 3 principle components of original extracted features generated from MRI patches which cantered on corresponding samples. (d) shows the scatter plots of first 3 principle components of semantic features generated from probability maps. The position of the viewer in (c) and (d) is azimuth = 10° and elevation = 20°.

*2.3 Auto-context model*

During inference, the classification forest is a collection of weak learners. In order to improve its performance, the auto-context model was used iteratively to leverage the surrounding information with respect to the object of interest. We used an initial classification random forest and anatomic features to create semantic information for all training patients, which were then used in combination with the original signatures to train an improved random forest. The process was repeated to train 4 classification random forests. In the segmentation stage, a new MR image can follow the same sequence of the auto-context method to obtain the segmentation. The influence of the auto-context method on segmentation is shown in our previous work [39].

*2.4 Evaluation*

We retrospectively analyzed MRI and CT data acquired during treatment planning for 14 patients who received cranial irradiation. The main patient selection criterion was that each MRI was acquired with the same 3D sequences and had fine spatial resolution, and the entire head was imaged. Standard T1-weighted MRI was captured using a GE MRI scanner with magnetization-prepared rapid gradient echo (MP-RAGE) sequence and 1.0×1.0×1.4 mm$^3$ voxel size (TR/TE: 950/13 ms, flip angle: 90°). CTs was captured with a Siemens CT scanner with 1.0×1.0×1.0 mm$^3$ voxel size with 120 kVp and 220 mAs. Bone, air and soft tissue were segmented on CT images and registered to MR images, which were used as ground truth. We used leave-one-out cross validation method to evaluate the proposed algorithm. To quantitatively evaluate of the performances, we calculated the DSC (air, bone and soft tissue) between the ground truth (CT segmentation) and the proposed method's segmentation[40].

$$\text{DSC} = \frac{2\times|X\cap Y|}{|X|+|Y|} \quad (2)$$

where *X* and *Y* are the ground truth contours and the contours obtained with the proposed method, respectively. We also calculated sensitivity and specificity using the overlapping ratio inside and outside the ground truth volume,

$$\text{Sensitivity} = \frac{|X\cap Y|}{|X|} \quad (3)$$

$$\text{Specificity} = \frac{|\bar{X}\cap\bar{Y}|}{|\bar{X}|} \quad (4)$$

where $\bar{X}$ and $\bar{Y}$ are the volumes outside the ground truth contours and contours obtained with the proposed method respectively.

To study the effectiveness of the proposed SCRF model, we ran the random forest (RF) method without an auto-context model and patch-based anatomical signatures and compared the resulting segmentation accuracy with the proposed method. Deep learning-based methods have been intensively studied in the last decade for various medical imaging applications. To compare the performance of SCRF method with deep learning-based methods, we also trained a well-established deep learning-based model, U-Net, for air, bone and soft tissue classification.

## 3. Results

*3.1 Comparison with random forest method*

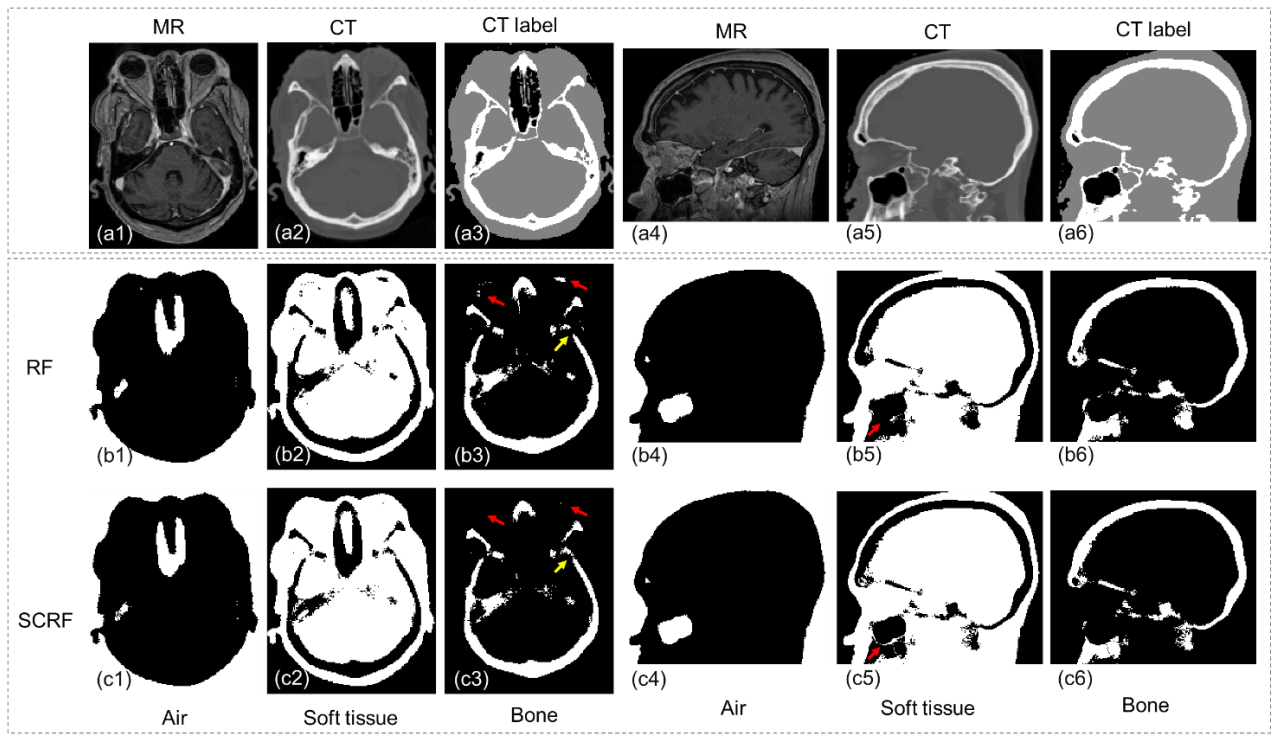

**Fig 5.** Qualitative comparison between RF and SCRF methods. (a1) and (a4) are MRI images shown in transverse and sagittal planes. (a2), (a5) and (a3), (a6) are corresponding CT images and CT labels (black for air, gray for soft-tissue, and white for bone) respectively. Row (b) and (c) are results generated with RF and SCRF respectively.

Fig. 5 shows the qualitative comparison between RF and SCRF methods. Both RF and SCRF generate air, soft tissue and bone classification similar to the classification obtained with CT images. However, the left lens was mislabeled as bone with RF method, as indicated by the red arrows in Fig. 5 (b3), and correctly classified as soft tissue with SCRF method. As shown by the red arrows in (b5) and (c5), the fine structure delineating maxillary sinus was better identified with SCRF.

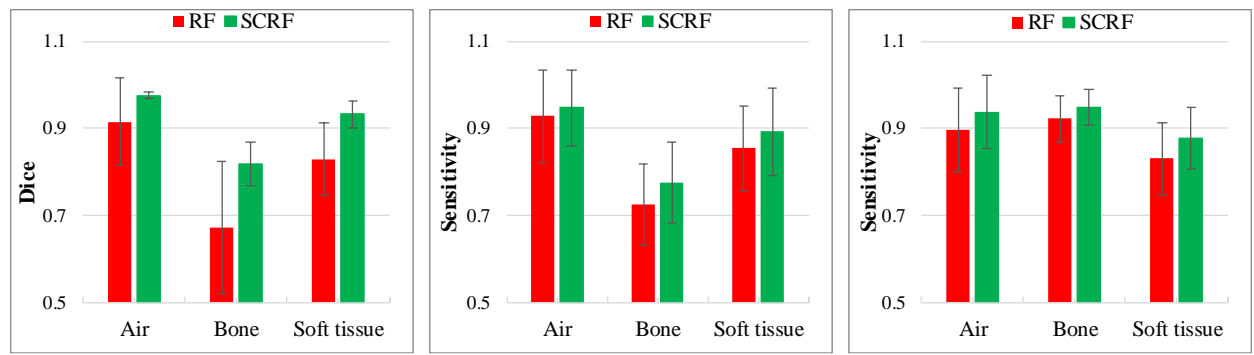

**Fig. 6** Quantitative comparison of RF and SCRF methods for DSC (left), sensitivity (middle), and specificity (right). Error bars show standard deviation.

Fig. 6 shows the quantitative comparisons of DSC, sensitivity and specificity between the two methods. With the integration of auto-context model and patch-based anatomical signature, the proposed SCRF method outperformed the RF method on all calculated metrics. The DSC on air, bone and soft tissue were 0.976±0.007, 0.819±0.050 and 0.932±0.031, compared to 0.916±0.099, 0.673±0.151 and 0.830±0.083 with RF. Sensitivities were 0.947, 0.775, 0.892 for air, bone and soft tissue with SCRF, and 0.928, 0.725, 0.854 with RF. Specificity for the three tissue types was also improved with proposed method, which were 0.896, 0.823 and 0.830 with RF, compared to 0.938, 0.948, 0.878 with proposed SCRF.

*3.2 Comparison with U-Net model*

We compared the performance of the proposed method against a well-established deep learning-based model, U-Net. As indicated by the red arrows in Fig. 7 (b2) and (c2), SCRF generated more accurate classification in challenging areas, such as the paranasal sinuses. U-Net failed to delineate the soft tissue around maxillary sinus (Fig. 7 (b5)), while the soft tissue structure was accurately identified with SCRF (Fig. 7 (c5)). SCRF also corrected the mislabeling produced by the U-Net method in Fig. 7 (b3). The performance improvement was further illustrated in the quantitative comparison. As shown in Fig. 8, U-Net obtained DSCs of 0.942, 0.791 and 0.917 for air, bone and soft tissue, and SCRF improved DSCs by 0.034, 0.028 and 0.016 respectively. Similarly, sensitivity calculated U-Net classification results were 0.927, 0.735 and 0.883 for air, bone and soft tissue, and increased by 0.020, 0.040 and 0.009 with SCRF. Specificity was improved by 0.019, 0.025 and 0.025 on the three tissue types with the proposed SCRF method.

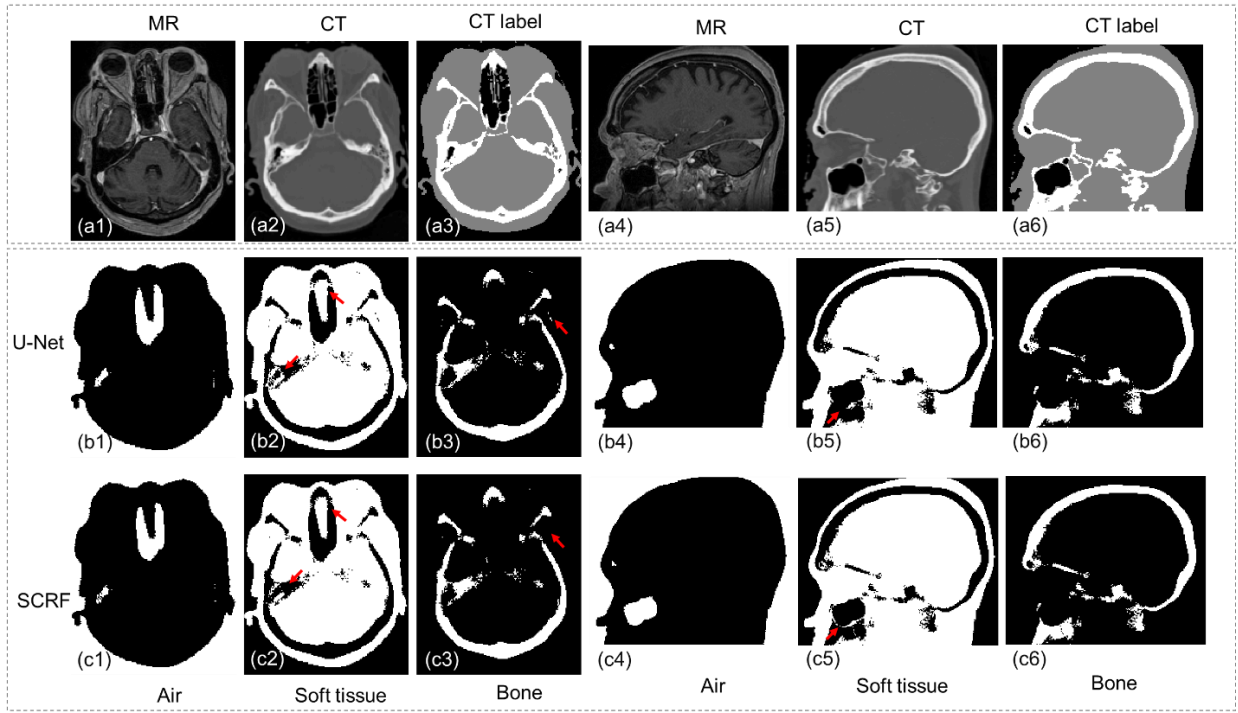

**Fig. 7** Qualitative comparison between U-Net and SCRF methods. (a1) and (a4) are MRI images shown in transverse and sagittal planes. (a2), (a5) and (a3), (a6) are corresponding CT images and CT labels (black for air, gray for soft-tissue, white for bone) respectively. Row (b) and (c) are results generated with RF and SCRF respectively.

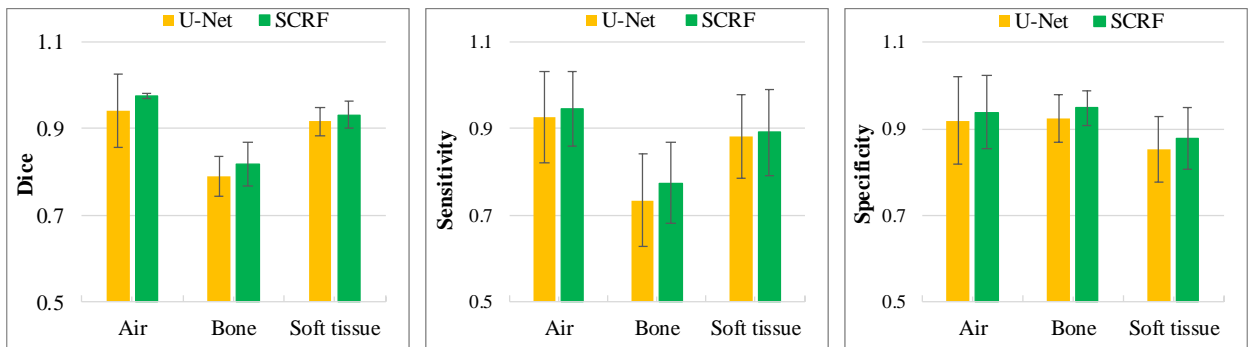

**Fig. 8** Quantitative comparison of U-Net and SCRF methods on DSC (left), sensitivity (middle) and specificity (right). Error bars show standard deviation.

## 4. Discussion

In this paper, we have investigated a learning-based approach to segment air, bone, and soft tissue from routine MRI. The novelty of our approach is the integration of an auto-context model, the contextual information from the predicted probability map and patch-based anatomical signature into a machine learning framework to iteratively segment MR images. In order to improve the random forest training efficiency, a feature selection mechanism was introduced to identify the more informative and salient features to serve as the anatomical signature of each voxel by minimizing the LASSO energy function. The selected features with higher discriminative power were used to train the random forest. Contrary to conventional segmentation methods, an auto-context model is used to incorporate contextual information from the previously discriminative probability maps in a random forest framework to provide an iterative refinement for the final segmentation, which significantly improves the segmentation accuracy. Experimental validation was performed to demonstrate its clinical feasibility and reliability. This segmentation technique could be a useful tool for MRI-based radiation treatment planning, attenuation correction for a hybrid PET/MRI scanner or MR-guided focused ultrasound surgery.

Special MR sequences, such as UTE [16, 17] and zero time echo (ZTE) [41, 42], have been investigated for bone detection and visualization, which were also employed on commercial PET/MRI systems for segmentation-based attenuation correction. However, the segmentation accuracy with conventional methods on those special sequences are usually limited due to the high level of noise and the presence of image artifacts. Juttukonda *et al.* derived intermediate images from UTE and Dixon images for bone and air segmentation, which obtained Dice coefficients of 0.75 and 0.60 for the two tissue types, respectively [43]. An *et al.* improved the UTE MR segmentation accuracy with a multiphase level-set algorithm [44]. The bone and air DSC obtained on 18F-FDG datasets were 0.83 and 0.62. Baran *et al.* built a UTE MR template that contained manual-delineated air/bone/soft tissue contours, and used Gaussian mixture models to fit UTE images, which obtained average Dice coefficients of 0.985 and 0.737 for air and bone [45]. The proposed SCRF method was implemented on MR images generated with routinely-acquired T1 sequences. Despite the limited bone-air contrast, our method demonstrated superior segmentation performance. Considering the valuable patient-specific bone extraction information provided by UTE and ZTE sequences, combining the superior detection capability of machine learning techniques with special bone-visualization sequences has the potential to generate promising results. In the future, we will explore the possibility of integrating the proposed method with UTE or ZTE MR sequences for better bone and air differentiation.

Several machine learning- and deep learning-based MR segmentation methods have been studied in the literature. Liu *et al.* trained a deep convolutional auto-encoder network for soft tissue, bone and air identification [22]. This deep learning-based method generated average Dice coefficients of 0.936 for soft tissue, 0.803 for bone, and 0.971 for air. Convolution encoder-decoder (CED) was also implemented on MR images acquired with both UTE and out-of-phase echo images for better bone and air differentiation, which generated mean Dice coefficient of 0.96, 0.88 and 0.76 for soft tissue, bone and air, respectively [46]. Comparing to those state-of-the-art methods, the proposed method demonstrates competitive segmentation performance.

We evaluate the performance of the proposed method on brain images, and demonstrated superior performances for bone and air segmentation even in challenging areas, such as paranasal sinuses. Bone and air segmentation is crucial for attenuation correction of brain PET images, and facilitates MR-guided focused ultrasound surgery. Bone segmentation in whole-body MR also have important clinical implications, such as musculoskeletal applications [47] and traumatic diagnoses [48, 49]. Different from brain MRI where the difficulty lies in the bone/air segmentation, challenges of bone segmentation on whole-body MRI are areas where spongy bone exists, such as the vertebra. The application of special MR sequences, such as UTE, is not clinically feasible for routine whole-body imaging due to the prolonged scanning time. For future work, we will modify the proposed method to include both cortical and spongy bone segmentation in the framework and implement it on whole-body MRI.

## 5. Conclusions

We proposed a machine learning-based automatic segmentation method that could identify soft tissue, bone and air on MR images. The proposed method has excellent segmentation accuracy, and the potential to be applied to attenuation correction for PET/MRI, MRI-only radiotherapy treatment planning and MR-guided focused ultrasound surgery.

## 6. Disclosures

No potential conflicts of interest relevant to this article exist

## 7. Acknowledgement

This research is supported in part by the National Cancer Institute of the National Institutes of Health under Award Number R01CA215718 (XY), and Dunwoody Golf Club Prostate Cancer Research Award (XY), a philanthropic award provided by the Winship Cancer Institute of Emory University.